\documentclass[twocolumn,superscriptaddress,longbibliography,aps,prb,preprintnumbers]{revtex4-2}

\usepackage[normalem]{ulem}
\usepackage{graphicx}
\usepackage{bm}
\usepackage{color}
\usepackage{epstopdf}
\usepackage{amsmath}
\usepackage{amssymb}
\usepackage{epstopdf}
\usepackage{lipsum}

\usepackage{gensymb}

\usepackage{multirow}

\usepackage{xfrac}

\usepackage{scrextend}

\usepackage[urlcolor=blue,colorlinks=true,citecolor=blue,linkcolor=blue,pdfstartview={FitH},bookmarks=false]{hyperref}

\graphicspath{{fig/}{./fig/}{.}}

\begin{document}

\title{
Phonon Weyl points and chiral edge modes \\ with unconventional Fermi arcs in NbSi$_{2}$
}

\author{Issam Mahraj}
\email[e-mail: ]{issam.mahraj@ifj.edu.pl}
\affiliation{Institute of Nuclear Physics, Polish Academy of Sciences, W. E. Radzikowskiego 152, PL-31342 Krak\'{o}w, Poland}

\author{Andrzej~Ptok}
\email[e-mail: ]{aptok@mmj.pl}
\affiliation{Institute of Nuclear Physics, Polish Academy of Sciences, W. E. Radzikowskiego 152, PL-31342 Krak\'{o}w, Poland}

\date{\today}

\begin{abstract}
NbSi$_{2}$ crystallizes in the P6$_{2}$22 symmetry, featuring chiral chains of Si atoms.
The absence of inversion symmetry, combined with its chiral structure, gives arise to unique physical properties.
The breaking of inversion symmetry leads to the emergence of Weyl points, while the chiral structure enables the formation of chiral edge modes.
As a result, NbSi$_{2}$ serves as an ideal platform for exploring the interplay between phonon Weyl points and chiral phonon edge modes.
For example, we identify the presence of a structure consisting of three Weyl points with a Chern number of $\mathcal{C} = +1$ around the $\bar{\text{K}}$ point.
These nodes form unconventional Fermi arcs connecting the $\bar{\Gamma}$ or $\bar{\text{K}}$ points, which mimic an effective Chern number of $\mathcal{C} = -2$.
\end{abstract}

\maketitle

%%%%%%%%%%%%%%%%%%%%%
%%%%%%%%%%%%%%%%%%%%%
%%%%%%%%%%%%%%%%%%%%%

\section{Introduction}

The recent discovery of three-dimensional (3D) topological materials has stimulated intensive investigations into compounds containing exotic quasiparticles. 
Topological insulators enable the realization of metallic surface states in the form of Dirac cones~\cite{zhang.liu.09,xia.qian.09,hsieh.xia.09,kuroda.arita.10}. 
Similarly, the nontrivial topology of Weyl semimetals manifests in surface states forming {\it Fermi arcs}~\cite{yan.felser.17,armitage.mele.18,lv.qian.21}. 
In these cases, the surface states connect two discrete, doubly degenerate points, which act as sources or sinks of Berry curvature in momentum space. 
Recently, these states have been experimentally observed in various materials~\cite{lv.xu.15,lv.weng.15,xu.belopolski.15,yang.liu.15,wang.zhang.16,xu.weng.16,xu.alidoust.17,yao.xu.19}.

However, topological features are not limited to electronic systems. Similar properties can be expected in phononic states~\cite{stenull.kane.16,liu.lian.17,jin.wang.18,liu.xu.19,liu.chen.20,li.wang.20,li.liu.21,liu.zou.22,wang.yang.22,liu.qian.20}.
In typical cases, topological properties are enforced by system symmetries~\cite{liu.qian.20,yu.zhang.22,zhu.wu.22,li.pan.24}. 
For example, Weyl points are expected in systems that lack inversion symmetry. 
In fact, the Weyl point with a Chern number of $\mathcal{C} = \pm 1$ is realized by two bands with linear dispersion around the touching point. 
Similarly, the touching of bands with higher-order dispersion can lead to the realization of multifold Weyl points with larger topological charges, such as $\mathcal{C} = \pm 2$~\cite{ding.zhou.22}, $\mathcal{C} = \pm 3$~\cite{wang.zhou.22,liu.chen.22}, or $\mathcal{C} = \pm 4$~\cite{liu.wang.21,fan.wan.23}. 
The realization of Weyl points in phononic states has been recently discussed in various systems~\cite{li.xie.18,xie.li.19,liu.hou.19,jin.chen.21,liu.li.21,liu.wang.21,peng.muakami.21,wang.zhou.22,liu.chen.22,ding.zhou.22,zhong.han.22,wang.zhou.22b,li.23,ding.xie.23,fan.wan.23,jin.hu.22}.

Another key aspect of topological properties is their association with chiral structures~\cite{chang.wieder.18}. 
Chirality, a feature in which an object cannot be superimposed onto its mirror image, strongly influences physical properties~\cite{barron.12,fecher.kubler.22,cheong.xu.22,bousquet.fava.25}. 
Chiral structures can serve as a source of chiral edge modes, which propagate unidirectionally around a closed loop. 
Such behavior has been reported in both electronic~\cite{chang.xu.17,tang.zhou.17,rao.li.19,li.xu.19,sanchez.belopolski.19,yuan.zhou.19,schroter.pei.19,takane.wang.19,yao.manna.20,sessi.fan.20,dai.li.22,dutta.ghosh.22,cochran.belopolski.23,mahraj.ptok.24} and phononic systems~\cite{zhang.song.18,miao.zhang.18,li.zhang.21,mahraj.ptok.24}.

{\it NbSi$_{2}$ physical properties.}--- 
Here, we focus on NbSi$_{2}$, which crystallizes in the P6$_{2}$22 symmetry~\cite{gottlieb.laborde.91,lasjaunias.laborde.93,sakamoto.fujii.05}. 
The chiral structure influences the physical properties of NbSi$_{2}$~\cite{onuki.nakamura.14}. 
For example, NbSi$_{2}$ exhibits the chirality-induced spin selectivity effect~\cite{shiota.inui.21,shishido.sakai.21}, in which a spin-polarized current is generated when an electric current passes through the system. 
Additionally, theoretical studies suggest that phonons play an important role in electronic transport~\cite{garcia.nenno.21}.

Moreover, the existence of screw rotational symmetry gives rise to Weyl points~\cite{tsirkin.souza.17,zhang.chan.18,gonzalez.tuiran.20}. Additionally, the chiral structure of NbSi$_{2}$ enforces the emergence of chiral edge modes.
Indeed, such chiral states are expected in the electronic band structure~\cite{zhang.shishidou.23}. 
These features make NbSi$_{2}$ an excellent platform for studying the interplay between Weyl nodes and chiral edge modes, especially given the recent advancements in high-resolution experimental techniques for phonon studies~\cite{miao.zhang.18,jin.hu.22,li.zhang.21}, which have made this field highly attractive.

The paper is organized as follows.
Details of the computational methods used are provided in Sec.~\ref{sec.method}.
In Sec.~\ref{sec.res}, we present and discuss our experimental and theoretical results.
Finally, a summary is given in Sec.~\ref{sec.sum}.

%%%%%%%%%%%%%%%%%%%%%
%%%%%%%%%%%%%%%%%%%%%
%%%%%%%%%%%%%%%%%%%%%

\begin{figure*}
\includegraphics[width=0.8\linewidth]{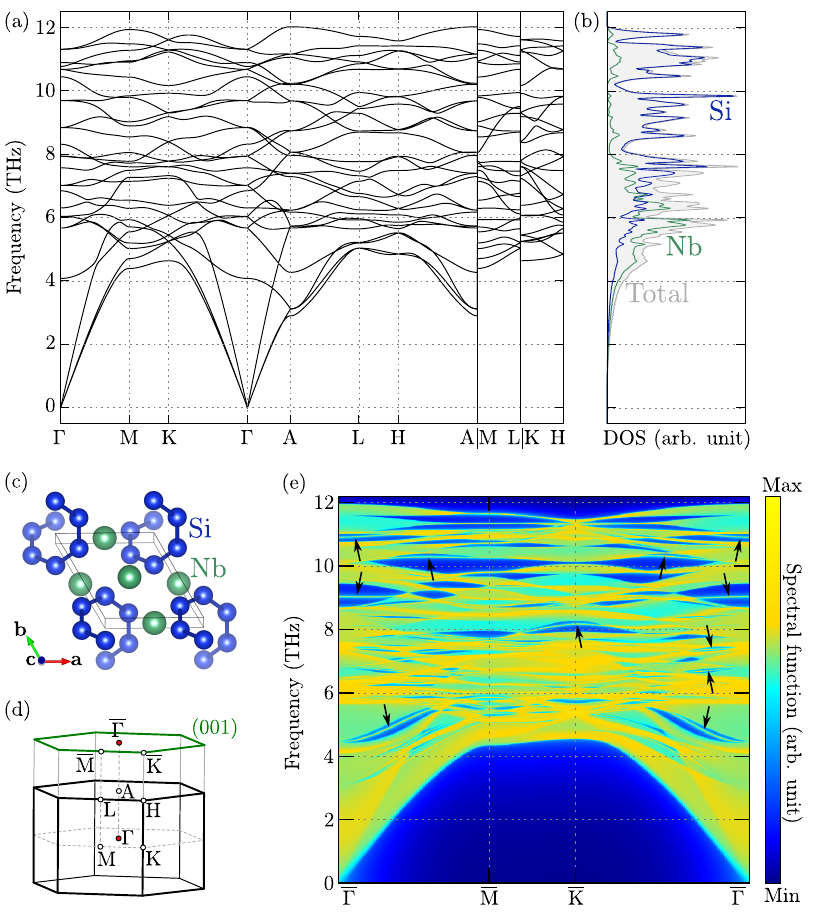}
\caption{
The phonon dispersion curves (a) and the density of states (b) for NbSi$_{2}$ with P6$_{2}$22 symmetry (c).
The corresponding bulk Brillouin zone and (001) surface Brillouin zone, along with their high-symmetry points, are presented in (d).
The phonon surface spectral function (e) exhibits several surface states (marked by red arrows).}
\label{fig.ph_band}
\end{figure*}

%%%%%%%%%%%%%%%%%%%%%
%%%%%%%%%%%%%%%%%%%%%
%%%%%%%%%%%%%%%%%%%%%

\section{Computational details}
\label{sec.method}

{\it Computational details.}---
The first-principles (DFT) calculations were performed using the projector augmented-wave (PAW) potentials~\cite{blochl.94} implemented in the Vienna Ab initio Simulation Package ({\sc Vasp}) code~\cite{kresse.hafner.94,kresse.furthmuller.96,kresse.joubert.99}.
Calculations were conducted within the generalized gradient approximation (GGA) using the Perdew, Burke, and Ernzerhof (PBE) parameterization~\cite{pardew.burke.96}.
The energy cutoff for the plane-wave expansion was set to $400$~eV.
Optimization of structural parameters (lattice constants and atomic positions) was performed in the conventional unit cell using a $12 \times 12 \times 8$ {\bf k}--point grid within the Monkhorst--Pack scheme~\cite{monkhorst.pack.76}.
The optimization loop was terminated when the energy difference fell below $10^{-6}$~eV for ionic degrees of freedom and $10^{-8}$~eV for electronic degrees of freedom.

Dynamic properties were calculated using the direct {\it Parlinski--Li--Kawazoe} method~\cite{parlinski.li.97}, implemented in the {\sc Phonopy} package~\cite{togo.chaput.23,togo.23}. 
Within this method, the interatomic force constants (IFCs) are obtained from the Hellmann--Feynman (HF) forces acting on atoms after individual atomic displacements inside the supercell.
We performed these calculations using a supercell consisting of $3 \times 3 \times 2$ unit cells.
During these calculations, a reduced $3 \times 3 \times 3$ ${\bm k}$-grid was used.
The obtained IFCs were then used to construct the tight-binding Hamiltonian.
The surface Green's function for a semi-infinite system~\cite{sancho.sancho.85} was computed using {\sc WannierTools}~\cite{wu.zhang.18}.

%%%%%%%%%%%%%%%%%%%%%
%%%%%%%%%%%%%%%%%%%%%
%%%%%%%%%%%%%%%%%%%%%

\section{Results and discussion}
\label{sec.res}

\subsection{Crystal structure}

The chiral transition-metal disilicide single crystal can crystallize into two {\it enantiomorphic} structures: a right-handed P6$_{2}$22 (space group No.~180) or a left-handed P6$_{4}22$ (space group No.~181).
These structures exhibit opposite chirality, determined by the sixfold screw axis aligned along the $c$ axis.
In the case of NbSi$_{2}$, the P6$_{2}$22 symmetry is preferred~\cite{sakamoto.fujii.05}.
Our first-principles calculations yield lattice parameters of $a = 4.8177$~\AA, $c = 6.6233$~\AA, which are in close agreement with the experimentally reported values $a = 4.819$~\AA, $c = 6.592$~\AA~\cite{kubiak.horyn.72}.
The primitive unit cell contains 9 atoms, i.e., 3 formula units.
The atomic positions are highly symmetric, with Nb occupying the Wyckoff position $3c$ $(\sfrac{1}{2},0,0)$ and Si positioned at $6i$ $(x_\text{Si}, 2 x_\text{Si}, 0)$, where $x_\text{Si} = 0.1591$.

\subsection{Lattice dynamic}

The crystal structure is dynamically stable, as evidenced by the absence of imaginary frequencies in the phonon dispersion curve [Fig.~\ref{fig.ph_band}(a)].
The acoustic branches near the $\Gamma$ point exhibit linear dispersion up to 3~THz.
The obtained phonon dispersion curves reproduce the previously reported result~\cite{emmanouilidou.mardanya.20} and are similar to those of isostructural NbGe$_{2}$ system~\cite{emmanouilidou.mardanya.20,garcia.nenno.21}.

Due to the significant mass contrast between Nb and Si, their vibrational modes are well separated in frequency: the heavy Nb atoms dominate the low-frequency range, whereas the lighter Si atoms primarily contribute to the high-frequency modes.
This separation is further confirmed by the calculated phonon density of states (DOS) [see Fig.~\ref{fig.ph_band}(b)].

The phonon modes at the $\Gamma$ point can be classified according to the irreducible representations of the space group $P6_{2}22$ as follows:
\begin{eqnarray}
\nonumber \Gamma_\text{acoustic} &=& A_{2} + E_{1}, \quad \text{and} \\
\nonumber \Gamma_\text{optic} &=& A_{1} + 2 A_{2} + 3 B_{1} + 2 B_{2} + 4 E_{1} + 4 E_{2}.
\end{eqnarray}
Among these, the $A_{2}$ and $E_{1}$ are infrared active, while the $A_{1}$, $E_{1}$, and $E_{2}$ modes are Raman active.
All active modes, except for the $A_{1}$, involve vibrations of all atoms in the crystal.
The $A_{1}$ modes, however, are exclusively associated with the vibrations of Si atoms, which form chiral chains [see Fig.~\ref{fig.ph_band}(c)]. 
The characteristic frequencies of the modes at the $\Gamma$ point, along with their corresponding irreducible representations, are listed in Table~\ref{tab.irr}.

%%%%%%%%%%%%%%%%%%%%%
%%%%%%%%%%%%%%%%%%%%%
%%%%%%%%%%%%%%%%%%%%%

\begin{table}[!t]
\caption{
\label{tab.irr}
Characteristic frequencies and symmetries of the phonon modes at the $\Gamma$ point for NbSi$_2$.}
\begin{ruledtabular}
\begin{tabular}{rrcc}
(THz) & (cm$^{-1}$) & Symmetry & Activity \\
\hline
4.214 & 140.56 & $B_{2}$ & -- \\
5.722 & 190.87 & $E_{1}$ & IR, Raman \\
5.939 & 198.10 & $E_{2}$ & Raman \\
6.409 & 213.78 & $B_{2}$ & -- \\
6.836 & 228.02 & $A_{2}$ & IR \\
6.875 & 229.33 & $E_{1}$ & -- \\
7.679 & 256.14 & $B_{2}$ & -- \\
7.811 & 260.55 & $E_{1}$ & IR, Raman \\
8.541 & 284.90 & $E_{2}$ & Raman \\
9.753 & 325.33 & $E_{2}$ & Raman \\
10.204 & 340.37 & $A_{1}$ & Raman \\
10.608 & 353.84 & $E_{1}$ & IR, Raman \\
10.757 & 358.81 & $B_{1}$ & -- \\
11.067 & 369.16 & $A_{2}$ & IR \\
11.603 & 387.03 & $E_{2}$ & -- \\
\end{tabular}
\end{ruledtabular}
\end{table}

\subsection{Phononic surface states}

The realization of the edge leads to the emergence of new modes not present in the bulk spectrum, i.e., surface states.
To investigate these modes, we compute the phonon surface spectral function for the (001) surface.
The resulting spectrum includes bulk states, originating from the projection of all states in the three-dimensional (3D) bulk Brillouin zone onto the two-dimensional (2D) surface Brillouin zone, as shown in Fig.~\ref{fig.ph_band}(d).
Accordingly, the bulk paths $\Gamma$--A, K--H, and M--L are projected onto the $\bar{\Gamma}$, $\bar{\text{K}}$, and $\bar{\text{M}}$ points, respectively.
Furthermore, due to the broken translational symmetry along the $z$-direction, surface states are clearly visible, as indicated by black arrows in Fig.~\ref{fig.ph_band}(e).
In practice, such surface states can be observed across the entire frequency range.

Phononic surface states correspond to vibrational modes associated with several atomic layers near the surface~\cite{li.wang.20}.
This raises an open question: how many layers actually participate in these vibrations?
To address this issue, let us briefly investigate the phononic surface states in a slab geometry.

\begin{figure}[!htp]
\includegraphics[width=\linewidth]{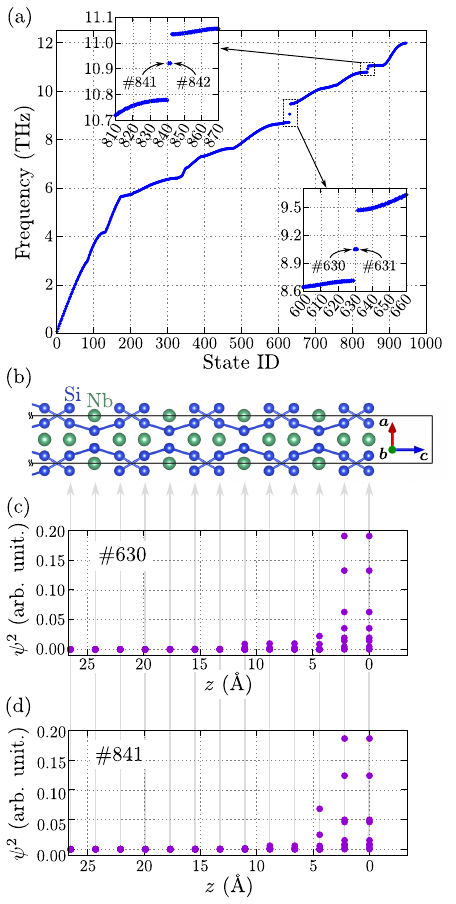}
\caption{Localization of the phononic surface states.
(a) The phonon spectrum of the NbSi$_{2}$ slab realzing the (001) surface, consisting of $35$ primitive unti cells (results show at the $\bar{\Gamma}$ point).
The edge of the system realizing the (001) surface is shown in (b).
Within the bulk gaps, two pairs of degenerate surface states can be found (states \#$630$/\#$631$ and \#$841$/\#$842$, indicated by arrows in the insets).
Localization of the \#$630$ and \#$841$ states is presented in (c) and (d), respectively.
Here, $\psi^{2}$ denotes the square of the polarization vector coefficients associated with atoms in a specific layer (marked by grey arrows along the panels).
}
\label{fig.local}
\end{figure}

Bulk vibrations are defined by the $\it{eigenproblem}$ of the dynamical matrix, which itself is obtained by a Fourier transformation of the IFCs from real space to momentum space, i.e., ${\bm R} \rightarrow {\bm k}$.
Thus, the vibration of each mode $\varepsilon$ at wave vector ${\bm k}$ is described by a polarization vector $e_{\varepsilon {\bm k} \alpha j}$, which characterizes the vibration of atom $j$ in the direction $\alpha \in {x, y, z}$.
This framework can be generalized to slab geometries, where the Fourier transformation is applied only in the directions parallel to the surface (e.g., $x$ and $y$ for the (001) surface).
In practice, such a transformation corresponds to a partial Fourier transformation: ${\bm R} = (R_x, R_y, R_z) \rightarrow (k_x, k_y, R_z)$, where the $z$ coordinate remains unchanged.
As a result, the wave vectors are now restricted to the plane that retains periodic boundary conditions (in this case, the $xy$ plane).
Simultaneously, the real-space coordinate $R_z$ remains unchanged and describes the atomic layers along the slab.
Consequently, the number of vibrational modes in the slab increases from $3N$ in the bulk (where $N$ is the number of atoms in the unit cell) to $3N \times M$, where $M$ is the number of unit cells in the slab.
Nevertheless, polarization vectors can still be used to analyze the vibrational modes.

Results obtained for the slab geometry with a (001) surface are presented in Fig.~\ref{fig.local}.
In our calculations, a slab consisting of 35 layers was used, yielding 945 modes for each wave vector.
Fig.~\ref{fig.local}(a) shows the spectrum at the $\bar{\Gamma}$ point, which directly corresponds to the $\Gamma$–A path in the bulk, as shown in Fig.~\ref{fig.ph_band}(d).
Along the $\Gamma$–A direction in the bulk phonon band structure, two band gaps are present [see Fig.~\ref{smfig.local} in the Supplemental Material (SM)~\footnote{See Supplemental Material at [URL will be inserted by publisher] for additional theoretical results.}].
Accordingly, two pairs of doubly degenerate edge modes can be clearly identified. These arise due to the breaking of periodic boundary conditions along the $z$ direction (i.e., perpendicular to the surface).
These edge states are highlighted in the two insets in Fig.~\ref{fig.local}(a), corresponding to modes \#$630$/\#$631$ and \#$841$/\#$842$.
For the aforementioned modes, we analyzed the polarization vectors in more detail.
Fig.~\ref{fig.local}(c) and~\ref{fig.local}(d) show the squared coefficients of the polarization vectors as a function of distance from the edge.
As seen, the atomic vibrations are predominantly localized within the three atomic layers closest to the edge (approximately $7$~\AA).
For deeper layers, the contributions to the polarization vectors become negligible.
These results can be compared with those obtained for electronic edge modes.
For example, in chiral EuPtSi, the electronic edge mode is mainly localized within two atomic layers at the edge (around $5$~\AA)~\cite{mahraj.ptok.24}.
Finally, it is worth noting that each mode in the identified pairs corresponds to vibrations localized at opposite edges of the slab.
At the $\bar{\Gamma}$ point, the atomic contributions to the vibrational modes are symmetric.
Moreover, the non-vanishing components of the polarization vectors are primarily associated with the Si atoms.

\begin{figure}[!htp]
\includegraphics[width=0.9\linewidth]{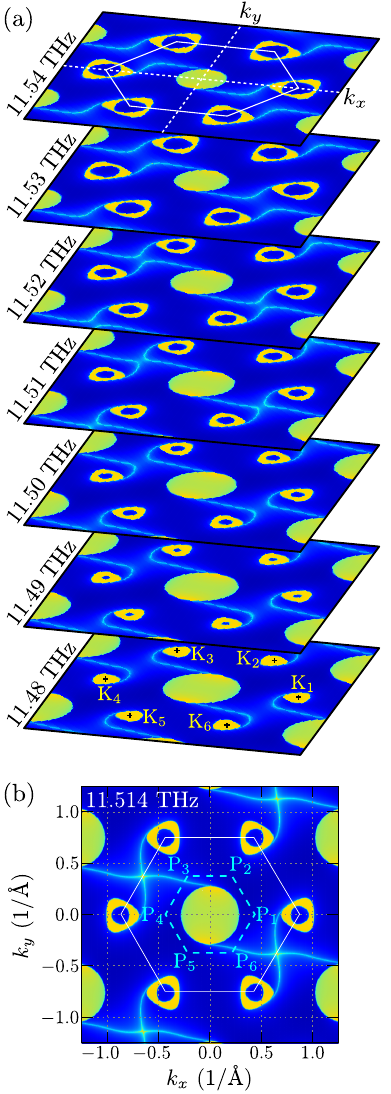}
\caption{
(a) Evolution of the chiral edge mode at several frequencies in the high-frequency range.
(b) Constant-frequency contour showing the touching of two Fermi arcs.
}
\label{fig.ph_band2a}
\end{figure}

\begin{figure}[!htp]
\includegraphics[width=\linewidth]{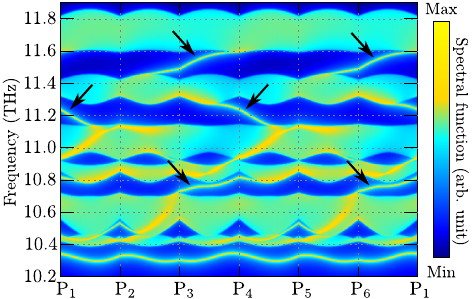}
\caption{
Spectral function along a closed contour around the $\overline{\Gamma}$ point, revealing chiral surface states (marked by red arrows).
The closed loop corresponds to the cyan dashed hexagon shown in Fig.~\ref{fig.ph_band2a}(b).
}
\label{fig.ph_band2b}
\end{figure}

\begin{figure*}
\includegraphics[width=\linewidth]{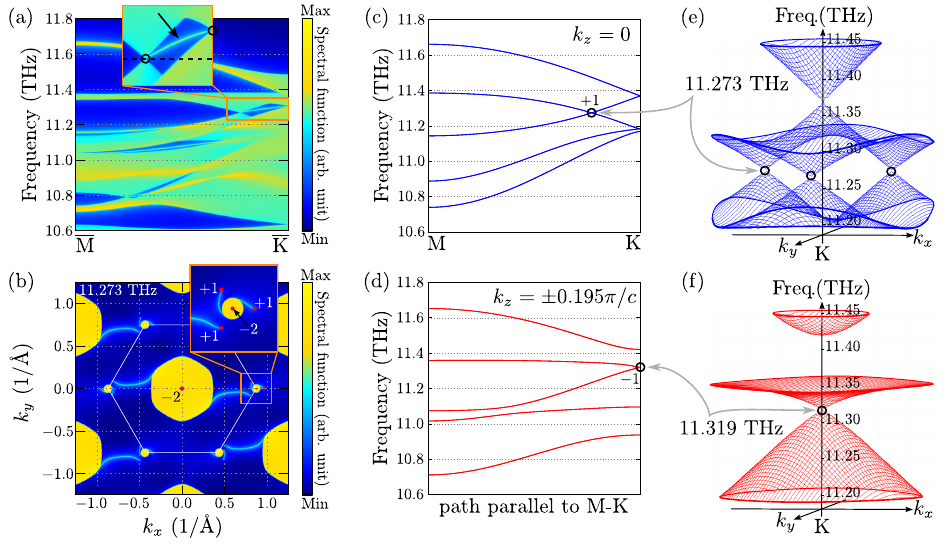}
\caption{
(a) The phonon spectral function along $\overline{\text{M}}$-$\overline{\text{K}}$ exhibits a topological surface state between Weyl points with opposite chiralities (marked by circles in the inset).
A constant-frequency cut at the lower Weyl point frequency (dashed line in (a)) reveals the Fermi arc connecting these Weyl points (b).
The $\overline{\Gamma}$ and $\overline{\text{K}}$ points exhibit a topological charge of $\mathcal{C} = -2$, while the three Weyl points around the $\overline{\text{K}}$ point have $\mathcal{C} = +1$.
Due to the bulk-boundary correspondence, this behavior originates from the bulk system properties.
The Weyl points with charge $\mathcal{C} = \pm 1$ are clearly visible in the phonon dispersion along the M-K path for $k_{z} = 0$ (c) and along a path parallel to the M-K path for $k_{z} = \pm 0.195 \pi / c$ (d).
For the Weyl points around K in the $k_{z} = 0$ plane, we observed three Dirac cone-like features.
In contrast, single Weyl points along the K-H path appear in the $k_{z} = \pm 0.195 \pi / c$ (f).}
\label{fig.ph_band3d}
\end{figure*}

\subsection{Chiral edge modes}

Several constant-frequency cross-sections are presented in Fig~\ref{fig.ph_band2a}(a).
Within the presented frequency range, we observed the evolution of surface states as the frequency increases. 
Regardless of the frequency, the sufrace states consistently connect one of the $\bar{\text{K}}$ points to the $\bar{\Gamma}$.
Notably, the number of surface states connected to a Weyl point is strongly related to its topological charge.
Thus, based on the form of the Fermi arcs, we identify the $\bar{\Gamma}$ point as a Weyl point with an effective Chern number $| \mathcal{C} | = 2$, while the $\bar{\text{K}}$ points have $| \mathcal{C} | = 1$.
Moreover, the observed surface states exhibit strong frequency and momentum dependence. 
As the frequency increases, the connection between the $\bar{\Gamma}$ point and the $\bar{\text{K}}$ points changes. 
Initially, we observe that the K$_{3}$ and K$_{6}$ points are connected to $\bar{\Gamma}$ within the first 2D Brillouin zone. 
However, above a certain frequency, the surface states switch, and $\bar{\Gamma}$ becomes connected to the K$_{1}$ and K$_{4}$ points. This transition occurs at a frequency of $11.514$~THz, where two edge modes intersect in the 2D Brillouin zone [see Fig.~\ref{fig.ph_band2a}(b)].
At the transition frequency, the surface states touch at a point on the edge of the 2D Brillouin zone. It is also important to note that this change in the shape of the surface states does not affect the Chern number values--i.e., the characteristic features around the high-symmetry points remain unchanged.

Chiral symmetry enables the realization of chiral edge modes.
In the surface spectral function, these modes appear as distinct modes propagation unidirectionally along a closed loop.
As a closed loop, we select a hexagonal path with vertices located at the midpoints of the K wavevectors [shown as a cyan dashed hexagon in Fig.~\ref{fig.ph_band2a}(b)].
The corresponding surface spectral function along this loop is presented in the inset of Fig.~\ref{fig.ph_band2b}.
Several groups of the bulk states appear as flat bands, and within the spectral gaps between them, we observe chiral modes propagating in different directions (indicated by black arrows).
Additionally, some surface states are hybridized with bulk states--for example, around $10.6$~THz.
Similar behavior is observed for a closed loop around the K point (see Fig.~\ref{smfig.loopk} in the SM~\cite{Note1}).

\subsection{Unconventional Fermi arcs and Weyl points}

We now focus on the high-frequency range of the surface spectral function. 
These frequencies are primarily associated with vibrations of the Si atoms within the chiral chains [cf. Fig.~\ref{fig.ph_band}(b)].
A zoomed-in view of the surface spectral function along the $\bar{\text{M}}$--$\bar{\text{K}}$ path is presented in Fig~\ref{fig.ph_band3d}(a). Several surface states cross the edge of the 2D surface Brillouin zone (e.g., around 11.1~THz and 11.3~THz). 
Moreover, some surface states appear between Weyl points [see inset in Fig.~\ref{fig.ph_band3d}(a)].
In this case, the Weyl points (marked by black circles) can exist beyond the high-symmetry points.
A constant-frequency cross-section at 11.273~THz (dashed line crossing the first Weyl point in the inset) reveals a complex structure of Fermi arcs [see Fig.~\ref{fig.ph_band3d}(b)]. 
As we can see, these Fermi arcs connect three Weyl points symmetrically located around the $\bar{\text{K}}$ points, extending to the $\bar{\Gamma}$ and $\bar{\text{K}}$ points. 
Thus, both the $\bar{\Gamma}$ and $\bar{\text{K}}$ points exhibit characteristic features associated with Weyl points carrying a topological charge of $| \mathcal{C} | = 2$. 
Similarly, the three Weyl points around the $\bar{\text{K}}$ point clearly exhibit features characteristic of a topological charge of $| \mathcal{C} | = 1$.

\begin{figure}[!t]
\includegraphics[width=\linewidth]{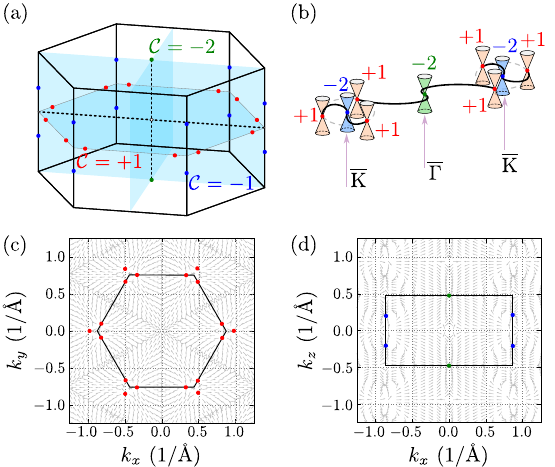}
\caption{
(a) Weyl points in the bulk Brillouin zone.
Weyl points with different Chern numbers $\mathcal{C}$ arise from the crossing between the 25th and 26th phonon bands.
Red, blue, and green points represent Weyl points (monopoles) with Chern numbers \mbox{$\mathcal{C} = +1$}, $-1$, and $-2$, respectively.
(b) Schematic representation of the relationship between Weyl points, Chern numbers, and edge modes (represented by black solid lines).
The Weyl point with \mbox{$\mathcal{C} = -2$} at the A point is projected onto the $\bar{\Gamma}$ point (green cone).
Similarly, two bulk points with \mbox{$\mathcal{C} = -1$} are projected onto the $\bar{\text{K}}$ point, creating an effective \mbox{$\mathcal{C} = -2$} point (blue cones), surrounded by three points with \mbox{$\mathcal{C} = +1$} (red cones).
Here, the effective absolute Chern number can be recognized as the number of edge modes touching a specific Weyl point (solid black lines).
(c) and (d) show the normalized Berry curvature vectors in the $k_{z} = 0$ and $k_{y} = 0$ planes, respectively.
Sources and sinks of the Berry curvature determine the positive and negative signs of the topological charge.}
\label{fig.topo}
\end{figure}

Such conventional analysis of the Fermi arcs provides information about the absolute value of the topological charge associated with the Weyl nodes.
In fact, the sign of the Chern number can also be determined from the Berry flux through a closed surface surrounding each node:
\begin{eqnarray}
\mathcal{C} = \frac{1}{2\pi} \oint \Omega \left( {\bm k} \right) \cdot d{\bm S} ,
\end{eqnarray}
where:
\begin{eqnarray}
\label{eq.berry_curv} \Omega \left( {\bm k} \right) = \nabla_{\bm k} \times A \left( {\bm k} \right)
\end{eqnarray}
is the Berry curvature, and
\begin{eqnarray}
A \left( {\bm k} \right) = \sum_{n=1}^{N} i \langle u_{n{\bm k}} | \nabla_{\bm k} | u_{n{\bm k}} \rangle
\end{eqnarray}
is the Berry connection.
The summation runs over all bands below the $N$th band, and $| u_{n\mathbf{k}} \rangle$ denotes the wavefunction of the $n$th band at momentum point $\mathbf{k}$.
Additionally, it is important to note that these Weyl points are associated with bulk band strucutre properties.
Indeed, an investigation of the bulk band structure along the M--K path reveals a band-touching point with a topological charge of $\mathcal{C} = +1$ [see Fig.~\ref{fig.ph_band3d}(c)].
Similarly, a touching point with $\mathcal{C} = -1$ is found along a path parallel to M--K at $k_{z} = \pm 0.195 \pi/c$ [see Fig.~\ref{fig.ph_band3d}(d)].
The corresponding 3D bulk band structures are shown in Fig.~\ref{fig.ph_band3d}(e) and Fig.~\ref{fig.ph_band3d}(f), respectively.
In the $xy$ plane ($k_{z} = 0$), the touching points are symmetrically distributed around the K point. 
In contrast, in the $k_{z} = \pm 0.195 \pi/c$ plane, only a single touching point is present. In both cases, linear-like dispersion is observed in the immediate vicinity of the touching points.

Let us briefly discuss the main features of the Weyl points realized in the high-frequency range (as discussed above).
These particular Weyl points arise from the touching of the 25th and 26th bulk phonon branches.
Their positions in the 3D Brillouin zone are shown in Fig.~\ref{fig.topo}(a). 
We identify three types of Weyl points: one with a topological charge of $\mathcal{C} = -2$ located at the A point, and several others with $\mathcal{C} = \pm 1$ situated around the K point.
Specifically, along the K--M path within the $k_z = 0$ plane, a Weyl point with $\mathcal{C} = +1$ is present. 
Similarly, along the K--H path at $k_z = \pm 0.195 \pi / c$, two Weyl points with $\mathcal{C} = -1$ are found. 
In total, the system hosts one Weyl point with $\mathcal{C} = -2$, four with $\mathcal{C} = -1$, and six with $\mathcal{C} = +1$ [represented by green, blue, and red points in Fig.~\ref{fig.topo}(a), respectively]. 
Despite this, the total topological charge over the entire Brillouin zone remains zero, as required by symmetry.
Additionally, the Weyl points along the K--H path in the 3D Brillouin zone all project onto the $\bar{\text{K}}$ point of the 2D surface Brillouin zone [which is schematically presented in Fig.~\ref{fig.topo}(b)].
As a result, the $\bar{\text{K}}$ point exhibits features characteristic of a $\mathcal{C} = -2$ Weyl point--namely, it is connected to two edge modes.

As is well known, a Weyl point can be treated as a monopole (source or sink) of Berry curvature in momentum space. 
To illustrate this, we calculate the Berry curvature using Eq.~(\ref{eq.berry_curv}).
The results obtained for the $k_z = 0$ and $k_y = 0$ planes are presented in Fig.~\ref{fig.topo}(c) and Fig.~\ref{fig.topo}(d), respectively.
As we can see, three sources ($\mathcal{C} > 0$) of the Berry curvature are clearly visible around each K point in the case of $k_{z} = 0$ plane [see Fig.~\ref{fig.topo}(c)].
Similarly, the sink of Berry curvature are visible in the $k_{y} = 0$ plane [see Fig.~\ref{fig.topo}(d)].

%%%%%%%%%%%%%%%%%%%%%
%%%%%%%%%%%%%%%%%%%%%
%%%%%%%%%%%%%%%%%%%%%

\section{Summary}
\label{sec.sum}

We investigate the NbSi$_{2}$ system in the context of topological phonons enforced by the system's symmetry~\cite{xu.vergniory.24}.
We present possible phonon chiral edge modes and their connection to multifold Weyl phonons.
The Weyl phonons are confirmed by Berry curvature calculations.
Additionally, we uncover the existence of unconventional Fermi arcs in the high-frequency range, associated with vibrations of the Si atoms in chiral chains.
Some of the Weyl points mimic a higher Chern number, which is reflected in the number of connected surface states.

NbSi$_{2}$ realizes the chiral noncentrosymmetric P6$_{2}$22 symmetry~\cite{sakamoto.fujii.05}.
However, the chirality-selected crystal growth technique using a laser-diode-heater floating zone method allows for the realization of either P6$_{2}$22 or P6$_{4}$22 structures~\cite{kousaka.sayo.23}.
The chirality of the obtained structure can be probed by single-crystal X-ray diffraction or spin polarization, originating from the chirality-induced spin selectivity effect~\cite{naaman.waldeck.12}.
A change in chirality from P6$_{2}$22 to P6$_{4}$22 corresponds to an inversion of the topological charges. Such physical properties make NbSi$_{2}$ an excellent platform for studying the interplay between multifold Weyl phonons and chiral edge modes.

\begin{acknowledgments}
Some figures in this work were rendered using {\sc Vesta}~\cite{momma.izumi.11} software.
This work was supported by the National Science Centre (NCN, Poland) under Project No.
2021/43/B/ST3/02166 (A.P.).
\end{acknowledgments}

%%%%%%%%%%%%%%%%%%%%%
%%%%%%%%%%%%%%%%%%%%%
%%%%%%%%%%%%%%%%%%%%%

%\nocite{*}
\bibliography{biblio.bib}

\clearpage
\newpage

\onecolumngrid

\begin{center}
  \textbf{\Large Supplemental Material}\\[.3cm]
  \textbf{\large Phonon Weyl points and chiral edge modes with unconventional Fermi arcs in NbSi$_{2}$}\\[.3cm]
  %%%%%%
  Issam Mahraj and Andrzej Ptok\\[.2cm]
  %%%%%%
  {\itshape
	Institute of Nuclear Physics, Polish Academy of Sciences, W. E. Radzikowskiego 152, PL-31342 Kraków, Poland
  }
  (Dated: \today)
\\[0.3cm]
\end{center}

\setcounter{equation}{0}
\renewcommand{\theequation}{S\arabic{equation}}
\setcounter{figure}{0}
\renewcommand{\thefigure}{S\arabic{figure}}
\setcounter{section}{0}
\renewcommand{\thesection}{S\arabic{section}}
\setcounter{table}{0}
\renewcommand{\thetable}{S\arabic{table}}
\setcounter{page}{1}

%%%%%%%%%%%%%%%%%%%%%%%%%%%%%%%%%%%
%%%%%%%%%%%%%%%%%%%%%%%%%%%%%%%%%%%
%%%%%%%%%%%%%%%%%%%%%%%%%%%%%%%%%%%

\begin{figure}[!h]
\includegraphics[width=0.6\linewidth]{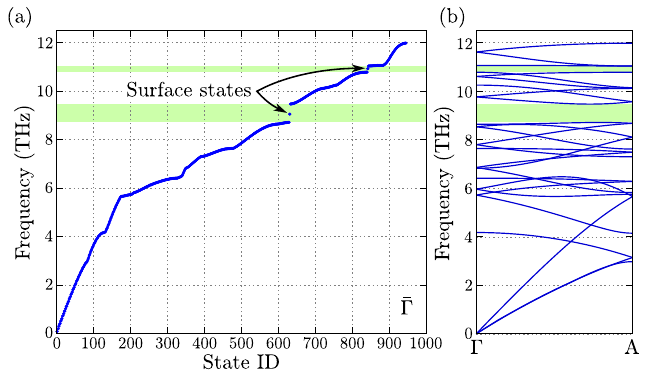}
\caption{
The phonon spectrum of the NbSi$_{2}$ slab realizing the (001) surface is shown in (a), consisting of 35 primitive unit cells.
The results are obtained at the $\bar{\Gamma}$ point.
This point can be compared with the bulk band structure along the $\Gamma$–A path, which represents the projection of the 3D bulk states onto the 2D surface Brillouin zone [as presented in Fig.~\ref{fig.ph_band}(d) in the main text].
Two band gaps are clearly visible in the bulk spectrum (marked by green areas), and the same gaps are also observed in the slab spectrum.
However, due to the presence of the edge, new states emerge inside these gaps, corresponding to surface states.
}
\label{smfig.local}
\end{figure}

\begin{figure}[!]
\includegraphics[width=\linewidth]{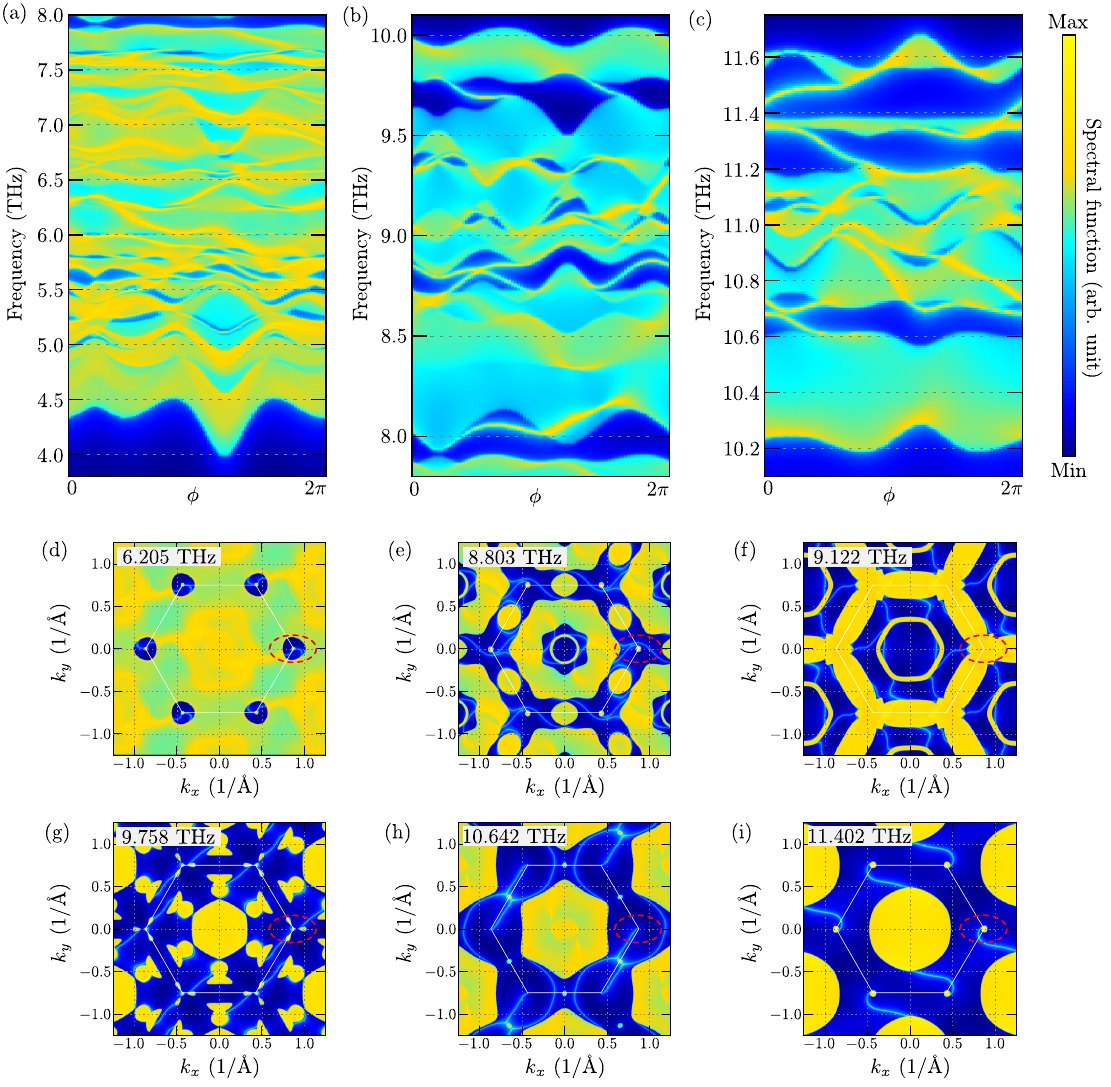}
\caption{
(a)-(c) Surface spectral function along a closed loop around the $\bar{\text{K}}$ point [loop marked by a red dashed ellipse in (d)].
(d)-(i) Constant frequency contours for several frequencies (as labeled).}
\label{smfig.loopk}
\end{figure}

In this Supplemental Material, we present additional results:
\begin{itemize}
\item Fig.~\ref{smfig.local} -- Correspondence between the phonon slab spectrum and the bulk phonon band structure.
\item Fig.~\ref{smfig.loopk} -- Surface spectral function along a closed loop around the  $\bar{\text{K}}$ point.
\end{itemize}

\vspace{3cm}

\end{document}